\documentclass[repr,laps,pra,twocolumn,floatfix,showpacs]{revtex4}
\usepackage[T1]{fontenc}
\usepackage[american]{babel}
\usepackage[colorlinks=true,citecolor=blue,linkcolor=blue,urlcolor=blue]{hyperref}
\usepackage{color}
\usepackage[dvips]{graphicx}
\usepackage{amsmath,amssymb,amsthm,amsfonts,mathptmx,pxfonts}
\usepackage{times}
\usepackage{subfigure}
\usepackage{psfrag}

\usepackage{mathpazo}
\usepackage[scaled]{helvet}
\usepackage{courier}
\normalfont
\include{Qcircuit}

\def\la{\langle}
\def\ra{\rangle}
\def\be{\begin{equation}}
\def\ee{\end{equation}}
\def\ba{\begin{eqnarray}}
\def\ea{\end{eqnarray}}
\newtheorem{definition}{HL}

\begin{document}

\title{Ultimate limits to quantum metrology and the meaning of the Heisenberg limit}
\author{Marcin Zwierz$^{1, 2}$}\email{zwierz.marcin@gmail.com}
\author{Carlos~A.~P\'{e}rez-Delgado$^{1, 3}$}
\author{Pieter Kok$^1$}
\affiliation{$^1$Department of Physics and Astronomy, University of
Sheffield, Hounsfield Road, Sheffield, S3 7RH, United Kingdom\\ $^2$ Centre for Quantum Computation and Communication Technology (Australian Research Council), Centre for Quantum Dynamics, Griffith University, Brisbane, QLD 4111, Australia\\ $^3$ Centre for Quantum Technologies, National University of Singapore,  3 Science Drive 2, Singapore 117543, Singapore}

\date{\today}

\begin{abstract}\noindent
For the last 20 years, the question of what are the fundamental capabilities of quantum precision measurements has sparked a lively debate throughout the scientific community. Typically, the ultimate limits in quantum metrology are associated with the notion of the Heisenberg limit expressed in terms of the physical resources used in the measurement procedure. Over the years, a variety of different physical resources were introduced, leading to a confusion about the meaning of the Heisenberg limit. Here, we review the mainstream definitions of the relevant resources and introduce the universal resource count, that is, the expectation value of the generator (above its ground state) of translations in the parameter we wish to estimate, that applies to all measurement strategies. This leads to the ultimate formulation of the Heisenberg limit for quantum metrology. We prove that the new limit holds for the generators of translations with an upper-bounded spectrum.
\end{abstract}

\maketitle

\section{Introduction}\noindent
Quantum metrology, or quantum parameter estimation theory, is an important branch of science that has received a lot of attention in recent years \cite{helstrom67,helstrom69,holevo11,caves81,braunstein94,braunstein96,lee02,giovannetti06,zwierz10,giovannetti11}. It studies high-precision measurements of physical parameters, such as phase, based on systems and physical evolutions that are governed by the principles of quantum mechanics. The main theoretical objective of this field is to establish the ultimate physical limits on the amount of information we can gain from a measurement \cite{helstrom67,helstrom69,holevo11,braunstein94,braunstein96}. From an experimental perspective, quantum-enhanced metrology promises many advances in science and technology, since an optimally designed quantum measurement procedure outperforms any classical procedure \cite{mitchell04,nagata07}. Furthermore, improved measurement techniques frequently lead not only to technological advancement, but also to a fundamentally deeper understanding of nature. The main figure of merit in the field of quantum metrology for both theorists and experimentalists is the precision with which the value of an unknown parameter can be estimated.

From this perspective, one of the most prominent concepts in quantum metrology is the Fisher information and the quantum Cram\'{e}r-Rao bound. The Fisher information $F(\phi)$ is a quantity that measures the amount of information about the parameter $\phi$ we wish to estimate revealed by the measurement procedure. Given the Fisher information, we can bound the minimal value of mean square error in the parameter with the quantum Cram\'{e}r-Rao bound. There exist two important regimes of the quantum Cram\'{e}r-Rao bound, the so-called shot-noise limit ({\sc snl}) that scales as $1/\sqrt{N}$ and the Heisenberg limit that scales as $1/N$, where $N$ is the resource count. The {\sc snl} is a limit attained by purely classical strategies (the term itself has its origin in quantum optics, where the detection of quanta of light is manifested as ``shots'' in a photon counter operating in Geiger mode \cite{giovannetti11}). The Heisenberg limit is imposed by the laws of quantum mechanics, namely, the generalized Heisenberg uncertainty relation, and for many years it was considered optimal and unbreakable. However, due to the unclear nature of the resource count, the optimality of the Heisenberg limit has recently been questioned \cite{luis05,boixo07,napolitano11}. 

A quantum measurement procedure can employ physical resources in a number of quantum states interacting with the measured system via various interaction mechanisms (e.g., linear, non-linear, or exponential) and operation strategies (either parallel or sequential). Therefore, in the literature a variety of meanings and definitions have been associated with $N$. In order to meaningfully compare different quantum and classical measurement procedures, it is essential to determine the relevant resources. The most versatile definition identifies $N$ with the number of times that the measured system is sampled. Although remarkably useful, this definition is not universal (e.g., when the number of quantum systems used in the measurement procedure is ill defined). In this paper, we introduce a universal definition of the physical resources which leads to the ultimate and optimal formulation of the Heisenberg limit for quantum metrology.

The paper is organized as follows. In \S~\ref{sec::hl}, we review various formulations of the Heisenberg limit for a fixed and limited amount of resources used in measurement procedures. In \S~\ref{sec::query}, we explain the concept of the query complexity, that is, the number of times the measured system is sampled, and demonstrate how it applies to a variety of well-known measurement procedures. In \S~\ref{sec::count}, we introduce a universal resource count for quantum metrology which leads to a new formulation of the Heisenberg limit. Finally, in \S~\ref{sec::conclusions}, we give some concluding remarks.

\section{Various formulations of the Heisenberg limit}\label{sec::hl}\noindent
In this section, we present a brief review of two definitions (and their interpretations) that are commonly associated with the term "Heisenberg limit" \cite{holland93,braunstein96,giovannetti11}. Not all of the interpretations are widely accepted. However, in our opinion this review properly reflects the present \textit{status quo} (i.e., the present confusion about the meaning) of the term ``Heisenberg limit'' in quantum metrology. What is the so-called Heisenberg limit or Heisenberg scaling?

In relation to the fundamental limitations of quantum metrology at least two interesting questions that are relevant for this field can be posed:
\begin{enumerate}
\item Given a fixed amount of resources, what is the best possible precision achievable in principle, that is, the precision that we aspire to reach?
\item (Given a physical setup), what is the precision that is actually obtained?
\end{enumerate}
The first question is usually answered by an appropriate expression of the quantum Cram\'{e}r-Rao bound leading to the Heisenberg scaling. The problem of attainability of the quantum Cram\'{e}r-Rao bound (and equivalently of the Heisenberg limit itself) is addressed by the second question. Given a physical setup, that is, a physical interaction between the probe and the sampled system, we minimize the error in a value of the parameter by employing optimal probe states and measurement observables. The mean-square error in parameter $\phi$ is then given by the error propagation formula:
\be
\Delta \phi = \frac{\Delta X}{| d\la X \ra/d\phi |}\, ,
\ee
where the average and standard deviation of an observable $X$ are calculated in an optimal state \cite{giovannetti06}. The derivative accounts for a possible change in units between the average value of the observable $X$ and parameter $\phi$. In this article, we are mainly concerned with finding an answer to the first question.

For many years, the notion of the Heisenberg limit $1/N$ has been linked with the best possible precision achievable in principle. This association was widely accepted and uncontroversial. The recent developments in quantum metrology seem to refute this claim, mainly as a consequence of the unclear nature of the resource count $N$ \cite{luis05,boixo07,napolitano11}. It is clear that the measurement procedure offering an arbitrarily high precision is physically unfeasible. In fact, it is possible to estimate the value of a parameter with perfect resolution only when there is some prior information available about the parameter, that is, when $\phi$ is \textit{a priori} limited to a particular range of values \cite{hall11}. In order to estimate the value of a truly continuous physical quantity \textit{distributed randomly} with an unbounded precision it is necessary to employ a probe with either an infinite number of constituents (e.g., a probe with an infinite number of photons) or an unbounded energy (e.g., an idealized continuous variable). Naturally, this approach is unphysical. In computer science the unbounded precision in representing the value of a continuous quantity is associated with an analog computer. It is well-known that the idealized analog computers are capable of solving problems that are intractable on digital computers, e.g., the {\sc np}-hard problems. The concept of the unbounded precision can also be linked with a digital machine (i.e., a Turing machine with an infinite memory capacity), wherein one can access infinitely many information carries (i.e., the classical or quantum bits) or execute the computation for infinitely many time steps. All those scenarios are clearly unphysical by being idealizations of a real-world phenomena. Both in physics and computer science the unbounded precision necessarily leads to the violations of elementary laws.

Historically the term ``Heisenberg limit'' was introduced by Holland and Burnett \cite{holland93}, who referred to the number-phase uncertainty relation in Heitler \cite{heitler54}. Hence, the very first formulation (recognized mainly by physicists) identifies $N$ with the number of physical systems in the probe, e.g., (average) photon number, which then can be easily related to the (average) energy of the probe. This formulation is clearly associated with the Heisenberg uncertainty relation and leads us to the first general definition of the Heisenberg limit.
\begin{definition}
The uncertainty in the value of an unknown parameter estimated with a single-shot Heisenberg-limited measurement procedure scales as
\be\label{uncer}
\Delta \phi \geq \frac{1}{2 \Delta \mathcal{H}}\, ,
\ee
where $\Delta \mathcal{H}$ is the standard deviation of the operator $\mathcal{H}$ that generates the translations of the probe state with the parameter $\phi$.
\end{definition}\noindent
To be more specific, this definition originates from the Mandelstam-Tamm type uncertainty relations $\Delta \phi \Delta \mathcal{H} \geq 1/2$ that are a manifestation of the generalized Heisenberg uncertainty relation with $\hbar = 1$ \cite{braunstein96,anadan90,boixo09}. For most of the measurement procedures, the standard deviation $\Delta \mathcal{H}$ can be easily expressed in terms of a variety of resources, including the number of quantum systems in the probe. As a consequence, the most well-known definition of the Heisenberg limit takes the following form.
\begin{definition}
The uncertainty in the value of an unknown parameter estimated with a single-shot Heisenberg-limited measurement procedure scales as
\be
\Delta \phi \geq \frac{1}{N}\, ,
\ee
where $N$ denotes the number (i.e., amount) of resources, typically, the (average) number of physical systems (e.g., photons) in the probe.
\end{definition}\noindent
Taking the (average) number of quantum systems in the probe as the fundamental resource count is appropriate and intuitively appealing in many important and practical measurement procedures. However, it has been clearly demonstrated that this expression of the Heisenberg limit is not universally valid \cite{luis05,boixo07,roy08}.

Given the abundance of different quantum and classical measurement procedures, a variety of resources were introduced, such as the already-mentioned number of quantum systems in the probe, the average energy of the probe, or the measurement time \cite{giovannetti11}. In the spirit of Landauer's famous conviction that information is inseparably connected with the underlying physical world, yet another formulation of the Heisenberg limit associates $N$ with the query complexity of a quantum network representing the measurement procedure \cite{giovannetti06,zwierz10}. From the ``physical perspective'', query complexity is equivalent to the number of fundamental physical interactions occurring between the probe and the sampled system. In most situations, the query complexity can be easily related to the number of physical systems in the probe, thus encompassing the earlier formulation. However, it is the formulation of the Heisenberg limit via the query complexity and not the number of quantum systems in the probe that properly captures the fundamental precision of most measurement procedures.

In the following section, we show that the Heisenberg limit is optimal with respect to the relevant resource count. To this end, we introduce the most general measurement procedure and then reduce it to a number of important measurement procedures, analyzing their performance with respect to the relevant resource count identified with the query complexity of a quantum network.

\section{Query complexity as the resource count}\label{sec::query}
\begin{figure}[!t]
\begin{flushright}
\includegraphics[scale=0.55]{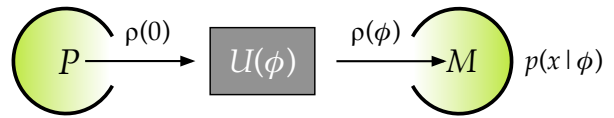}
\caption{The general parameter estimation procedure involving state preparation $P$, evolution $U(\phi)$, and generalized measurement $M$ with outcomes $x$, which produces a probability distribution $p(x|\phi)$. \label{PE}}
\end{flushright}
\end{figure}

\noindent
Let us briefly recall the structure of the most general measurement procedure (see Fig.~\ref{PE}). Any estimation procedure can be divided into three basic steps:
\begin{enumerate}
\item  a probe system (sensitive to the parameter we wish to estimate) is prepared in an initial quantum state $\rho(0)$. The probe consists of a fixed and limited number of physical systems that can be either well-defined or known on average; 
\item the state of the probe system is evolved to a state $\rho(\phi)$ by $U(\phi)=\exp(-i\phi\mathcal{H})$. The Hermitian operator $\mathcal{H}$ is the generator of translations in $\phi$, the parameter we wish to estimate. Physically, this corresponds to the interaction between the probe system and the sampled system; 
\item the probe system is subjected to a generalized measurement $M$, described by a positive operator valued measure ({\sc povm}), and the value of $\phi$ is extrapolated using data processing of the measurement results \cite{giovannetti11}.
\end{enumerate}
Note that in general, any process (including any quantum estimation procedure) can be represented as an instance of a quantum computation involving state preparation, evolution, and measurement. Due to this universality, any estimation procedure can be written as a quantum network. It is, therefore, intuitively clear that the query complexity of quantum networks should offer a valuable insight into the inner workings and the performance of measurement procedures based on those networks.

\begin{figure}[!t]
\begin{center}
\includegraphics[scale=1.2]{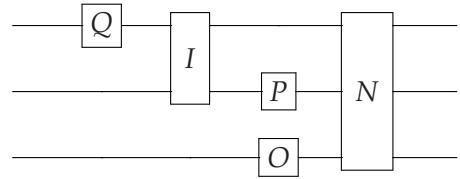}
\end{center}
\caption{An example of a quantum network consisting of single ($Q$, $P$, $O$) and many-body ($I$, $N$) quantum gates. \label{network}}
\end{figure}

Quantum networks arise naturally in the circuit model of quantum computation. A quantum network can be represented or intuitively understood as a series of geometric figures. These figures consist of horizontal wires representing qubits (or in general any quantum systems) and quantum gates. The gates perform simple computational tasks on the information carried by the quantum systems. Typically, a quantum network involves many quantum systems and many quantum gates (see Fig.~\ref{network}). We represent a quantum gate as a function $f(x_{1}, \ldots, x_{N})$ with a fixed number of input parameters and a fixed number of output parameters. Here, we employ a special type of the quantum gate called a black box or a quantum oracle. A black box is a unitary operator defined by its action on quantum systems whose internal workings are usually unknown. Crucially, a black box acts in a consistent way on a well-defined set of quantum systems. As a result, we can associate with any quantum network (acting on a well-defined number of quantum systems) the concept of the query complexity representing the number of times the black box appears in this network. Mathematically, a black box is a function than can be univariate or multi-variate. When the function is multi-variate, e.g., a bi-variate function of two arguments, then a query to the black box must consist of two input parameters. This reasoning extends to many-body black-box operators that in the setting of quantum metrology describe the basic interactions between the probe and the sampled system. 

Giovannetti, Lloyd, and Maccone were the first to show that there exists a fundamental connection between the concept of the query complexity and the field of quantum metrology \cite{giovannetti06}. The key insight of their result was that the precision of any (non-entangling) measurement procedure should be given in terms of the number of black-box interactions, that is, the query complexity \cite{giovannetti06}. The versatility of this approach was also emphasized by Braunstein, who advocated that the language of black boxes (each introducing the same unknown parameter) is general and can be applied to a rich class of measurement strategies \cite{braunstein06}. Similarly, van Dam \textit{et al}. addressed the problem of estimating the phase given $N$ copies of the (black-box) phase rotation gate \cite{vandam07}. The connection between the query complexity and the precision in estimating the value of the parameter was clarified and formalized for higher-order (or many-body) generators of translations in the parameter \cite{zwierz10}. While query complexity had been used before in the context of quantum metrology \cite{giovannetti06}, this paper is the first to analyze the most general interaction that governs the evolution of the probe system in this light. 

Here, we focus on the query complexity of the quantum network that governs the evolution stage of the measurement procedure. Let us consider a completely general quantum network that encompasses all possible measurement strategies. The most general quantum evolution acting on the probe system generated by the operator $\mathcal{H}$ is represented by the unitary transformation $U(\phi)$, which can be graphically represented by
\begin{equation}\label{net}
		\Qcircuit @C=0.66em @R=0.8em @!R  {
  & \gate{{V_0}} 	&  \gate{{O(\phi)}}	& \gate{{V_1}} 	& \gate{{O(\phi)}} 	& \qw & \ldots & &\gate{{O(\phi)}} & \gate{{V_Q}} & \qw } \, .
\end{equation}
This general interaction consists of $Q$ applications of a black-box operator $O(\phi) = \exp(-i \phi H)$ (where $H$ is a positive Hermitian generator), interspersed with $Q+1$ arbitrary unitary gates $V_j$. The arbitrary unitary gates $V_j$ together with ancillary systems may be used to introduce adaptive (feed-forward) strategies to the estimation procedure. For a general interaction $U(\phi)$, we can use an argument by Giovannetti, Lloyd, and Maccone \cite{giovannetti06} to show that the generator of $U(\phi)$ is given by
\be
\mathcal{H} = i \left(\frac{\partial U(\phi)}{\partial \phi}\right) U^{\dagger}(\phi)  = \sum_{j = 1}^{Q}  A_{j} \, ,
\ee
where
\ba
A_{j} &=& i V_{Q}\, O(\phi)\, \ldots\, V_{j+1}\, O(\phi)\, V_{j}\, \frac{\partial O(\phi)}{\partial \phi}\, V^{\dag}_{j}\, O^{\dag}(\phi)\, V^{\dag}_{j+1}\, \ldots \nonumber \\
&&\ldots\, O^{\dag}(\phi)\, V^{\dag}_{Q}\, .
\ea
Therefore, the number of terms in the generator $\mathcal{H}$ is \textit{always} equal to the number of black-box operators appearing in the quantum network, that is, $Q$. Also note that the spectrum of the generator of a black-box operator $O(\phi)$ is unchanged by the unitary operators $V_{j}$. As a result, the spectrum of the generator $\mathcal{H}$ depends solely on the spectrum of the black-box operator. In the following we reduce this most general quantum network to specific quantum networks corresponding to some of the most well-known measurement procedures and relate their performance to the query complexity of underlying quantum networks.

\begin{figure}[!t]
\includegraphics[scale=0.8]{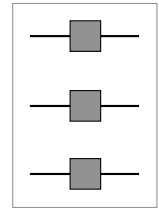}
\caption{The evolution stage of a linear (non-entangling) measurement procedure involving $N = 3$ quantum systems. The query complexity is linear in the number of quantum systems, that is, $Q = N$. \label{GLM}}
\end{figure}

\subsection{Linear measurement procedure}\noindent
Let us begin with a standard (non-entangling) linear procedure. Those kinds of procedures were described by Giovannetti, Lloyd, and Maccone \cite{giovannetti06}. In the linear case, the same parameter $\phi$ is applied independently on $N$ indistinguishable quantum systems (see Fig.~\ref{GLM}). Here, each fundamental physical interaction of the form: $O_{j}(\phi) = \exp(-i \phi H_{j})$, where $j$ denotes the quantum system, represents a single query. Consequently, the joint generator of the evolution operator $U(\phi)$ can be written as a sum of commuting generators $H_{j}$, that is,
\be
\mathcal{H}_{GLM} = \sum_{j=1}^{N} H_{j}\, .
\ee
Since the query complexity corresponds to the number of terms in the joint generator $\mathcal{H}_{GLM}$, we have $Q=N$. The fundamental precision then scales as \cite{giovannetti06}
\be
\Delta \phi \geq \frac{c_{1}}{Q} = \frac{c_{1}}{N} = O(N^{-1})\, ,
\ee
with $c_{1} = 1/(\lambda_{\rm max} - \lambda_{\rm min})$ with $\lambda_{\rm max}$ and $\lambda_{\rm min}$ being the maximal and minimal eigenvalues of $H$. Typically, $c_{1}$ is a constant of order one. This bound is saturated by the following family of maximally entangled states of the probe \cite{giovannetti06}:
\be
|\psi\ra = \frac{1}{\sqrt{2}}\left(|h_{\rm max}\ra + e^{i \varphi} |h_{\rm min}\ra\right)\, , \label{sat}
\ee
where $|h_{\rm max}\ra = |\lambda_{\rm max}\ra_{1} \cdots |\lambda_{\rm max}\ra_{N}$ and $|h_{\rm min}\ra = |\lambda_{\rm min}\ra_{1} \cdots |\lambda_{\rm min}\ra_{N}$ are the eigenstates corresponding to the maximal and minimal eigenvalues of $\mathcal{H}$, respectively (with $|\lambda_{\rm max}\ra$ and $|\lambda_{\rm min}\ra$ being the eigenstates corresponding to the maximal and minimal eigenvalues of $H$, respectively). In the setting of quantum interferometry, the formal analog of the maximally entangled state given by Eq.~(\ref{sat}) is a {\sc NOON} state \cite{lee02,kok10}:
\be
|\psi\ra = \frac{1}{\sqrt{2}} \left( |N,0\ra + |0,N\ra \right)\, ,
\ee
in which $N$ photons are propagating along the first or the second optical path of the Mach-Zehnder interferometer \cite{giovannetti11}.

\begin{figure}[!t]
\includegraphics[scale=0.8]{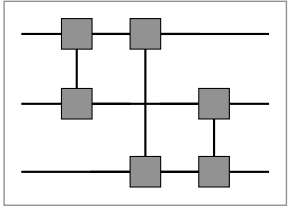}
\caption{The evolution stage of a two-body (entangling) measurement procedure involving $N = 3$ quantum systems. The query complexity scales quadratically with the number of quantum systems, that is, $Q = \frac{1}{2}N(N-1)$. \label{BFCG}}
\end{figure}

\subsection{Many-body measurement procedure}\noindent
The evolution operator $U(\phi)$ can also act on the constituents of the probe in a non-trivial (non-linear) way. In the case of the so-called many-body or higher-order measurement procedures introduced by Boixo, Flammia, Caves, and Geremia \cite{boixo07}, the fundamental interactions are applied to multiple quantum systems. For example, a two-body joint generator $\mathcal{H}_{BFCG}$ then takes two quantum systems as an input:
\be
\mathcal{H}_{BFCG} = \sum_{l=1}^{N} \sum_{j=1}^{l} H_{j} \otimes H_{l}\, ,
\ee
and is depicted in Fig.~\ref{BFCG} (with $N=3$) as a collection of bi-variate black-box operators $O_{jl}(\phi) = \exp(-i \phi H_{j} \otimes H_{l})$ (where $j$ and $l$ label the quantum systems) that constitute fundamental two-body interactions. In this case a single query is necessarily applied to two input quantum systems. From this it follows that given a probe consisting of $N$ quantum systems a total number of queries is the number of possible pairs from a set of size $N$, that is, $Q = \left({N \atop 2}\right) = \frac{1}{2}N(N-1)$. It is, therefore, clear that the number of queries $Q$ \textit{is not always} identical to the number of physical systems $N$ in the probe. The fundamental precision of a two-body measurement procedure expressed in terms of the query complexity is given by
\be
\Delta \phi \geq \frac{c_{2}}{Q} = c_{2} \left({N \atop 2}\right)^{-1} = O(N^{-2})\, ,
\ee
where $c_{2} = 1/(\lambda^{2}_{\rm max} - \lambda^{2}_{\rm min})$, with $\lambda_{\rm max}$ and $\lambda_{\rm min}$ being the maximal and minimal eigenvalues of $H$ \cite{boixo07}. Typically, $c_{2}$ is a constant of order one. Note that the error in $\phi$ is linear in the query complexity $Q$. At the same time, $Q$ has a superlinear scaling with $N$ attributed to the entangling power that the evolution operator has over the constituents of the probe. This approach naturally extends to generators of any degree $k \leq N/2$. A $k$-body measurement procedure offers the following scaling \cite{boixo07}
\be
\Delta \phi \geq \frac{c_{k}}{Q} = c_{k} \left({N \atop k}\right)^{-1} = O\left(N^{-k}/k!\right)\, ,
\ee
where $c_{k} = 1/(\lambda^{k}_{\rm max} - \lambda^{k}_{\rm min})$. This bound is also saturated by the family of maximally entangled state of the probe given by Eq.~(\ref{sat}). 

\begin{figure}[!t]
\includegraphics[scale=0.8]{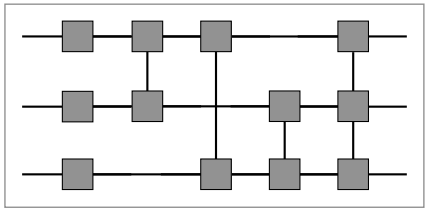}
\caption{The evolution stage of an exponential (highly-entangling) measurement procedure involving $N = 3$ quantum systems. The query complexity scales exponentially with the number of quantum systems, that is, $Q = 2^{N}-1$. \label{RB}}
\end{figure}

\subsection{Exponential measurement procedure}\noindent
The finite number of quantum systems in the probe imposes restrictions on the dimensionality of the probe's Hilbert space. By exploiting the whole Hilbert space, Roy and Braunstein introduced the exponential measurement procedure \cite{roy08}. In Fig.~\ref{RB}, we present this procedure translated to the language of the query complexity. Here, the query complexity scales exponentially with the number of quantum systems, that is, $Q = 2^{N}-1$. The fundamental precision of the exponential measurement procedure expressed in terms of the query complexity is given by
\be
\Delta \phi \geq \frac{c_{e}}{Q}  = O(2^{-N})\, , \label{exp}
\ee
where $c_{e}$ is a constant of order one \cite{roy08}. Interestingly, this bound is saturated by separable states due to the fact that all necessary entanglement is being generated at the evolution stage of the measurement procedure. 

Kitaev's famous phase estimation algorithm on discrete quantum systems, e.g., qubits, is another example of a measurement procedure which offers precision that scales exponentially with the number of employed qubits \cite{kitaev,cleve}. However, as in the previous case, this procedure requires an exponential number of fundamental unitary evolution gates, that is, black boxes. As a result, the precision is again bounded by Eq.~(\ref{exp}).

We emphasize that the bounds expressed in terms of the number of queries $Q$ are saturated by the appropriate optimal states, that is,
\be
\Delta \phi  \simeq \frac{1}{Q}\, ,
\ee
for all measurement procedure with well-defined $Q$. Therefore, the relevant resource count can be identified with the query complexity offering the linear scaling of the root-mean-square error in $\phi$. This indicates that the query complexity \textit{may} be considered a good resource count. However, does this mean that the query complexity \textit{is} the proper physical resource count for quantum metrology? We give the answer to this question in the next section.  

\subsection{Query complexity and the shot-noise limit}\noindent
First, however, we need to remark that the concept of the query complexity is much more natural for quantum-enhanced metrology and generally does not apply to the classically limited procedures. The main reason for this is the problematic nature of the {\sc snl} itself.

In the case of a linear strategy, the shot-noise-limited (classical) measurement procedure consists of $N$ independent measurement repetitions. each involving a single black-box interaction offering the following scaling:
\be
\Delta \phi  \geq \frac{c_{1}}{\sqrt{Q}} = \frac{c_{1}}{\sqrt{N}}\, \label{linear}
\ee
as prescribed by the quantum Cram\'{e}r-Rao bound. The same scaling is found by calculating the standard deviation of the generator $\mathcal{H}$ in the separable state of $N$ quantum systems, that is, $\Delta \mathcal{H} = \sqrt{\sum_{j=1}^{N}\Delta^{2} H_{j}} = c_{1} \sqrt{N} = c_{1} \sqrt{Q}$. However, these approaches fail in the case of many-body (non-linear) measurement strategies. For example, the shot-noise-limited $k$-body measurement procedure can be defined (in analogy to the linear case) as $N$ independent measurement repetitions, each involving a single $k$-variate black-box interaction yielding the same scaling as the scaling given by Eq.~(\ref{linear}). On the other hand, Boixo \textit{et al.} \cite{boixo08} more formally derived an $O(N^{k-1/2})$ scaling offered by a $k$-body measurement procedure fed with the separable state of $N$ quantum systems. Here, because of the Big $O$ notation it is impossible to calculate a well-defined number of queries. Also, the two approaches no longer predict the same scaling with $N$ [As a side remark: in the case of the exponential measurement procedure it is even nonsensical to consider the {\sc snl} as there is no single, basic black box that can be repeated $N$ times and the separable state is the optimal state for this procedure]. As a consequence, the concept of the query complexity does not apply to the classical procedures. It seems that the main difficulty in extending the language of the query complexity to the classical domain lies in the fact that it is unclear what is the universal definition of the shot-noise limit that would apply to all types of measurement strategies. If the {\sc snl} is a limit obtained in a procedure consisting of $N$ independent repetitions of a basic black box, then it yields a trivial bound, that is, it always scales as $1/\sqrt{Q}$, with $Q=N$ [the advantage of this approach is a well-defined query complexity]. If the {\sc snl} is a limit obtained in a procedure that employs a separable state, then for the $k$-body procedure the query complexity is ill defined.

What is the universal definition of the {\sc snl} limit that can be consistently applied to any estimation procedure and can it be expressed in terms of the query complexity? We leave these as open questions.

\section{Universal resource count}\label{sec::count}\noindent
The concept of the query complexity proved extremely useful in setting fundamental limits on the capabilities of a variety of measurement procedures and relating these to the number of employed quantum systems. We demonstrated that the query complexity can be used to meaningfully compare the precision offered by these measurement procedures. Moreover, we showed that the error in the parameter scales linearly with $Q$ for a number of important measurement procedures. Also, it is straightforward to tie the query complexity with the number of physical systems in the probe. This implies that the query complexity is a good resource count. Unfortunately, the query complexity can be ill defined. Some measurement procedures have an ill defined number of physical systems in the probe, and as a result, the exact number of queries is unknown. For example, for an optical measurement procedure employing coherent states, the number of photons in the probe is known only on average and there is no such quantity as an average number of queries. Therefore, we need to find a universal resource count that can deal with these cases properly.

Any universal resource count for quantum metrology should fulfill some basic requirements. First, for any measurement procedure the minimal uncertainty in the value of the parameter must scale linearly with a universal resource count. Second, in order to find this resource count, and by implication a universal formulation of the Heisenberg limit, we cannot refer to a specific physical implementation. Instead, we should derive the fundamental resource count from the general description of a measurement procedure. Finally, for purely physical reasons a good candidate for a universal resource count should also relate in a straightforward manner to some quantum-mechanical observable such as the number of physical systems in the probe or its (average) energy. 

Traditionally, the Heisenberg limit on the interferometric precision of estimating a phase $\phi$ is generally understood as the following scaling relation:
\be
\Delta \phi \geq \frac{c}{\la N \ra}\, ,
\ee
where $c$ is a constant of order one and $\la N \ra$ is the average of the number operator $N = a^{\dagger} a$ which generates the phase shift, that is, the total number of photons in the probe \cite{hall11}. This is a well-established relation \cite{hall11,giovannetti11arXiv,tsang11} and its achievability has been recently proved \cite{hayashi11}. For general quantum parameter estimation the number operator is replaced by the operator $\mathcal{H}$ which generates the translations in the parameter.  

Given the traditional formulation of the Heisenberg limit (and keeping in mind the above requirements), it is natural to formalize the universal resource count for quantum metrology as the expectation value of the generator of translations in the parameter $\la \mathcal{H} \ra$. Note that it is necessary to set the minimal eigenvalue (the ground-state eigenvalue) of the generator $\mathcal{H}$ to zero. The necessity of rescaling the value of the resource count stems from the fact that when $\mathcal{H}$ corresponds to a proper Hamiltonian, the origin of the energy scale has no physical meaning and as a consequence, we must fix the scale such that the ground state has zero energy. Therefore, the universal resource count is given by the expectation value of the generator $\mathcal{H}$ above its ground state denoted as $|\la\mathcal{H}\ra|$. We can also formulate our resource count without any shift in terms of $\la\mathcal{H}-h_{\rm min}I\ra$, where $h_{\rm min}$ is the minimal eigenvalue of $\mathcal{H}$ and $I$ is the identity operator. When $\mathcal{H}$ does not have a minimum eigenvalue, as in the case of position or momentum operators, the only possible values for $\la\mathcal{H}\ra$ are \emph{relative} position and momentum.

As we show in the following subsections, the proposed resource count fulfils all the requirements of a universal resource count and applies to any conceivable measurement procedure, even when apparently no query complexity can be defined. We argue that $|\la\mathcal{H}\ra|$ is a more fundamental resource count than $Q$, since it can deal with these cases as well. Nevertheless, whenever the query complexity exists, it is desirable to find an exact relation between the universal resource count and $Q$ (and by implication $N$).

\subsection{Standard deviation of the generator $\mathcal{H}$}\noindent
It can be argued that the standard deviation of the generator $\mathcal{H}$ can also serve as a universal resource count \cite{brody96,brody97}. Indeed, in the next subsection we show that $\Delta\mathcal{H}$ is related to the query complexity, and by implication to the number of quantum systems in the probe. However, $|\la \mathcal{H} \ra|$ is the \textit{only} moment that fits the right category, given the question of how many resources are required to attain a certain precision. Resources are ``a certain amount of something''. Thus, when dealing with probabilistic situations the physical amount is given by the first moment, while the higher-order moments describe the shape of the distribution. Also, it is important to note that the first moment represents a fundamental conserved quantity \cite{margolus11arXiv}. This distinction captures a physical intuition that goes beyond the pure mathematics of quantum metrology.

\begin{figure}[!t]
\centering
\includegraphics[scale=0.2]{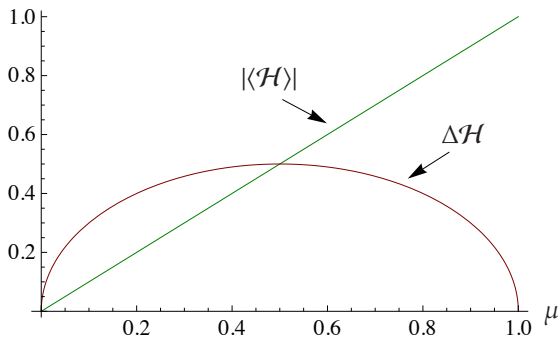}
\caption{The expectation value of the generator $\mathcal{H}$ above the ground state versus the standard deviation of the generator $\mathcal{H}$ as a function of $\mu$ [with $\| \mathcal{H} \| = 1$ a.u.]. \label{exp_vs_stan}}
\end{figure}

In order to quantitatively capture the distinction between the expectation value of the generator $\mathcal{H}$ above the ground state and the standard deviation of the generator $\mathcal{H}$, let us consider the (important for quantum metrology) family of pure superpositions given by
\be\label{eq:super}
|\psi\ra = \sqrt{\mu} |h_{\rm max}\ra + \sqrt{1-\mu} \, e^{i \varphi} |h_{\rm min}\ra\, ,
\ee 
where $|h_{\rm max}\ra$ and $|h_{\rm min}\ra$ are the eigenstates corresponding to the maximal and minimal eigenvalues of $\mathcal{H}$, respectively. The expectation value of the generator $\mathcal{H}$ in the above state can be written as 
\be
\la \mathcal{H} \ra = \mu h_{\rm max} + (1 - \mu) h_{\rm min}\, , 
\ee
where $h_{\rm max}$ and $h_{\rm min}$ are the maximal and minimal eigenvalues of $\mathcal{H}$, respectively. The expectation value of $\mathcal{H}$ above its ground state and the standard deviation of $\mathcal{H}$ are then given by
\ba
|\la\mathcal{H}\ra| \equiv \la\mathcal{H}-h_{\rm min}I\ra &=& \mu (h_{\rm max} - h_{\rm min}) = \mu \|\mathcal{H}\|\, , \nonumber \\
\Delta \mathcal{H} &=& \sqrt{\mu (1 - \mu)} (h_{\rm max} - h_{\rm min}) \nonumber \\
&=& \sqrt{\mu (1 - \mu)} \|\mathcal{H}\| \, , \nonumber
\ea 
where $\| \mathcal{H} \|$ is the semi-norm of the generator $\mathcal{H}$. In Fig.~\ref{exp_vs_stan}, we depict the above relations [with $\| \mathcal{H} \| = 1$ a.u.]. Note that both quantities are upper bounded by $\| \mathcal{H} \|$: 
\be
|\la\mathcal{H}\ra| \leq \| \mathcal{H} \| \ \ , \ \ \Delta \mathcal{H} \leq \frac{\| \mathcal{H} \|}{2}\, \nonumber
\ee
and coincide for $\mu = \frac{1}{2}$. Alas, even for as specific a family of superpositions as the one given by Eq.~(\ref{eq:super}), no definite relation between $|\la \mathcal{H} \ra|$ and $\Delta \mathcal{H}$ can be established. 

The importance of the standard deviation $\Delta \mathcal{H}$ stems from the fact that this quantity provides the achievable bounds in quantum metrology, that is, it gives the answer to the second question posed in \S~\ref{sec::hl}. On the other hand, the expectation value $|\la \mathcal{H} \ra|$ being a universal physical resource count provides the bound on the best possible precision achievable in principle (see the first question posed in \S~\ref{sec::hl}).

In the next subsection, we derive a lower bound on the error in the parameter $\phi$ in terms of a new resource count $|\la\mathcal{H}\ra|$. We also present the exact relations between the query complexity, the expectation value of $\mathcal{H}$ above its ground state, and the standard deviation of $\mathcal{H}$ for the most relevant measurement strategies.

\subsection{New formulation of the Heisenberg limit}\label{sec::formulation}\noindent
Having established the proper resource count, we present a new formulation of the Heisenberg limit with respect to this resource count. We consider here the most general quantum measurement procedure corresponding to the unitary transformation $U(\phi)$ presented in \S~\ref{sec::query}. Since the Heisenberg limit should refer to the {\em optimal} scaling behavior of the error with the resource count (it is a bound that we aspire to reach, not the actual achievable bound in any given experimental setup), we can restrict our discussion to the case of optimal states for quantum metrology, such as the textbook case of {\sc NOON} states \cite{lee02,kok10}. The optimal states in quantum metrology are the families of balanced superpositions of the eigenvectors $|h_{\rm max}\ra$ and $|h_{\rm min}\ra$ of $\mathcal{H}$, that is, the state given by Eq.~(\ref{eq:super}) with $\mu = \frac{1}{2}$. For the optimal states we have the property that
\be\label{eq:semi-norm}
2 \Delta\mathcal{H} = 2 |\la\mathcal{H}\ra| = \| \mathcal{H} \| = h_{\rm max} - h_{\rm min}\, . 
\ee
Combining this with Eq.~(\ref{uncer}), the error in parameter $\phi$ in a single-shot experiment is then given by
\be\label{HL}
\Delta\phi \geq \frac{1}{2 |\la\mathcal{H}\ra|}\, .
\ee
This inequality holds (and is tight) for the quantum states that are \textit{optimal} for quantum metrology, and is therefore an expression for the minimum error $\Delta\phi$ that can be achieved in an optimal measurement. While the derivation is mathematically valid only for generators with an upper-bounded spectrum, physically this bound will generally be satisfied since we can always truncate the Hilbert space at sufficiently high energy states.  Therefore, given a system evolution described by $U(\phi) = \exp(-i\phi \mathcal{H})$, for any (numerical) value for $|\la\mathcal{H}\ra|$ (i.e., the resource amount), the best attainable precision for a measurement of $\phi$ is bounded by Eq.~(\ref{HL}). This is a universal formulation of the Heisenberg limit. The new bound is just as tight as the bound provided by the standard deviation of $\mathcal{H}$, and whenever the latter exists both are identical.

In the following, we present the exact relations between $\Delta\mathcal{H}$, $|\la\mathcal{H}\ra|$ and the query complexity $Q$ that applies to the most relevant measurement strategies presented in \S~\ref{sec::query}. Given the arguments about the spectrum of the generator $\mathcal{H}$ for the most general quantum measurement procedure (see \S~\ref{sec::query}), for the optimal states the maximal and minimal eigenvalues $\mathcal{H}$ are given by
\be\label{eq:eigen}
h_{\rm max} = Q \lambda^{k}_{\rm max} \quad\text{and}\quad h_{\rm min} = Q \lambda^{k}_{\rm min} \, ,
\ee
where $\lambda_{\rm max}$ and $\lambda_{\rm min}$ are the maximal and minimal eigenvalues of $H$, respectively. The power $k$ denotes the order of the black-box interaction: with $k = 1$ representing a linear (non-entangling) black-box interaction. Since the exponential measurement procedures have a more complex structure, their corresponding $h_{\rm max}$ and $h_{\rm min}$ have a more compound form as well:
\ba
h_{\rm max} = \sum_{j=1}^{N} Q_{j} \lambda^{j}_{\rm max} \quad \text{and} \quad h_{\rm min} = \sum_{j=1}^{N} Q_{j} \lambda^{j}_{\rm min} \, 
\ea
with
\ba
Q_{j} = \left({N \atop j}\right) \quad \text{and} \quad Q = \sum_{j=1}^{N} Q_{j} = 2^{N} - 1\, .
\ea
Therefore, for the sake of clarity, we present our final result using bounds given by Eq.~(\ref{eq:eigen}). Combing these with Eq.~(\ref{eq:semi-norm}), we have
\be
2 \Delta\mathcal{H} = 2 |\la\mathcal{H}\ra| = \| \mathcal{H} \| = Q (\lambda^{k}_{\rm max} - \lambda^{k}_{\rm min})\, .
\ee
This leads to a bound on the error in parameter $\phi$, expressed in terms of the query complexity
\be
\Delta\phi \geq \frac{1}{2 |\la\mathcal{H}\ra|} = \frac{c_{k}}{Q}\, 
\ee
with $c_{k} = 1/(\lambda^{k}_{\rm max} - \lambda^{k}_{\rm min})$ being the inverse of the largest gap in the spectrum of the generator $\mathcal{H}$. Regardless of whether a well-defined number of queries exists or not, the first bound always holds. Given a particular measurement procedure, it is then straightforward to express the error in the parameter in terms of the number of quantum systems in the probe via the query complexity. For example, in the standard cases of linear (non-entangling) evolutions, the query complexity (and by implication $2 |\la\mathcal{H}\ra|$) reduces to the number of physical systems in the probe.

Finally, we note that our result applies to both parallel and sequential (or ``multi-round") measurement strategies, always giving a well-defined bound that can be expressed in terms of either $|\la\mathcal{H}\ra|$ or $Q$ (whenever the query complexity can be defined, it is always a finite number). For a parallel procedure (i.e., no sequential repetitions of the evolution gate $U(\phi)$ are allowed), the query complexity is limited by the dimensionality of probe's Hilbert space, thus, $Q \leq 2^{N} -1$. In the case of a sequential strategy (i.e., sequential repetitions of the evolution gate $U(\phi)$ possibly interspersed with some arbitrary unitaries are allowed), the total number of queries is limited by the dimensionality of the probe's Hilbert space and the measurement time $T$ (i.e., the number of repetitions), thus, $Q \leq T \times (2^{N} -1)$. In general, the sequential strategies may be looked upon as a way of estimating a value of a rescaled parameter. In these cases, one can argue that a value of the parameter $\theta = T \phi$ is being estimated rather than $\phi$. Nevertheless, it is worth emphasizing that the sequential strategies are more general and often offer some advantages over the parallel strategies, e.g.,  through the simplicity of their implementation \cite{giovannetti11}.

\section{Conclusions}\label{sec::conclusions}\noindent
Proper resource accounting is crucial when investigating the precision of various quantum measurement strategies and formulating the ultimate limits in quantum metrology. In this work, we applied the concept of the query complexity representing the number of times the measured system is sampled to a variety of well-known measurement procedures analyzing their performance. This leads to a universal definition of the physical resources (formalized as the expectation value of the generator of $\mathcal{H}$ above its ground state) and the ultimate formulation of the Heisenberg limit for quantum metrology. The new bound holds only for optimal states and is just as tight as the bound provided by the standard deviation of $\mathcal{H}$ (whenever the latter exists, both are identical).

\begin{acknowledgments}\noindent
This work was supported by the White Rose Foundation and the ARC Centre of Excellence CE110001027.
\end{acknowledgments}


\end{document}